\documentclass[twocolumn,superscriptaddress]{revtex4}

\usepackage{graphicx}
\usepackage{amsmath,amssymb}
\usepackage{subfigure}

\newcommand\mat{\mathbf}
\renewcommand\vec{\mathbf}

\newcommand\etal{\textit{et~al.}}
\newcommand\set[1]{\lbrace#1\rbrace}
\newcommand\kin{k^\textrm{in}}
\newcommand\kout{k^\textrm{out}}

\begin{document}

\title{Community structure in directed networks}

\author{E. A. Leicht}
\affiliation{Department of Physics, University of Michigan, Ann Arbor,
MI 48109, U.S.A.}
\author{M. E. J. Newman}
\affiliation{Department of Physics, University of Michigan, Ann Arbor,
MI 48109, U.S.A.}
\affiliation{Center for the Study of Complex Systems, University of
  Michigan, Ann Arbor, MI 48109, U.S.A.}

\begin{abstract}
  We consider the problem of finding communities or modules in directed
  networks.  The most common approach to this problem in the previous
  literature has been simply to ignore edge direction and apply methods
  developed for community discovery in undirected networks, but this
  approach discards potentially useful information contained in the edge
  directions.  Here we show how the widely used benefit function known as
  modularity can be generalized in a principled fashion to incorporate the
  information contained in edge directions.  This in turn allows us to find
  communities by maximizing the modularity over possible divisions of a
  network, which we do using an algorithm based on the eigenvectors of the
  corresponding modularity matrix.  This method is shown to give
  demonstrably better results than previous methods on a variety of test
  networks, both real and computer-generated.
\end{abstract}


\maketitle

At the most fundamental level a network consists of a set of nodes or
vertices connected in pairs by lines or edges, but many variations and
extensions are possible, including networks with directed edges, weighted
edges, labels on nodes or edges, and others.  This flexible structure lends
itself to the modeling of a wide array of complex systems and networks
have, as a result, attracted considerable attention in the recent physics
literature~\cite{AB02,DM2002,Newman2003c,Boccaletti2006}.

Many networks are found to display ``community structure,'' dividing
naturally into communities or modules with dense connections within
communities but sparser connections between them.  Communities have proven
to be of interest both in their own right as functional building blocks
within networks and for the insights they offer into the dynamics or modes
of formation of networks, and a large volume of research has been devoted
to the development of algorithmic tools for discovering
communities---see~\cite{DDDA2005} for a review.  Nearly all of these
methods, however, have one thing in common: they are intended for the
analysis of undirected network data.  Many of the networks that we would
like to study are directed, including the world wide web, food webs, many
biological networks, and even some social networks.  The commonest approach
to detecting communities in directed networks has been simply to ignore the
edge directions and apply algorithms designed for undirected networks.
This works reasonably well in some cases, although in others it does not,
as we will see in this paper.  Even in the cases where it works, however,
it is clear that in discarding the directions of edges we are throwing away
a good deal of information about our network's structure, information that,
at least in principle, could allow us to make a more accurate determination
of the communities.

Several previous studies, including our own, have touched on this problem
in the context of other analyses of directed network
data~\cite{NL2007,GSA07,ADFG2007,RB2007}, but they have typically not
tackled the community structure problem directly.  In this paper we propose
a method for the discovery of communities in directed networks that makes
explicit use of the information contained in edge directions.  The method
we propose is an extension of the well established modularity optimization
method for undirected networks~\cite{Newman2004a}, a method that has been
shown to be both computationally efficient and highly effective in
practical applications~\cite{DDDA2005}.

The premise of the modularity optimization method is that a good division
of a network into communities will give high values of the benefit
function~$Q$, called the modularity, defined by~\cite{NG2004}
\begin{align}
Q &= \mbox{(fraction of edges within communities)} \nonumber\\
  &\qquad{} - \mbox{(expected fraction of such edges)}.
\label{eq:modDeff}
\end{align}
Large positive values of the modularity indicate when a \emph{statistically
  surprising} fraction of the edges in a network fall within the chosen
communities; it tells us when there are more edges within communities than
we would expect on the basis of chance.

The expected fraction of edges is typically evaluated within the so-called
configuration model, a random graph conditioned on the degree sequence of
the original network, in which the probability of an edge between two
vertices~$i$ and~$j$ is $k_ik_j/2m$, where $k_i$ is the degree of
vertex~$i$ and $m$ is the total number of edges in the network.  The
modularity can then be written
\begin{equation}
Q = {1\over2m} \sum_{ij} \biggl[ A_{ij} - {k_ik_j\over2m} \biggr]
    \delta_{c_i,c_j},
\label{eq:moddef2}
\end{equation}
where $A_{ij}$ is an element of the adjacency matrix, $\delta_{ij}$ is the
Kronecker delta symbol, and $c_i$ is the label of the community to which
vertex~$i$ is assigned.  Then one maximizes~$Q$ over possible divisions of
the network into communities, the maximum being taken as the best estimate
of the true communities in the network.  Neither the size nor the number of
communities need be fixed; both can be varied freely in our attempt to find
the maximum.

In practice, the exhaustive optimization of modularity is computationally
hard, known to be NP-complete over the set of all graphs of a given
size~\cite{Brandes07}, so practical methods based on modularity
optimization make use of approximate optimization schemes such as greedy
algorithms, simulated annealing, spectral methods, and others.

Now consider a directed network.  In searching for communities in such a
network we again look for divisions of the network in which there are more
edges within communities than we expect on the basis of chance, but we now
take edge direction into account.  The crucial point to notice is that the
expected positions of edges in the network depend on their direction.
Consider two vertices, A~and~B.  Vertex~A has high out-degree but low
in-degree while vertex~B has the reverse situation.  This means that a
given edge is more likely to run from~A to B than \textit{vice versa},
simply because there are more ways it can fall in the first direction than
in the second.  Hence if we \emph{observe} in our real network that there
is an edge from~B to~A, it should be considered a bigger surprise than an
edge from~A to~B and hence should make a bigger contribution to the
modularity, since modularity should be high for statistically surprising
configurations.

We put these insights to work as follows.  Given the joint in/out-degree
sequence of our directed network, we can create a directed equivalent of
the configuration model, which will have an edge from vertex~$j$ to
vertex~$i$ with probability $\kin_i\kout_j/m$, where $\kin_i$ and $\kout_j$
are the in- and out-degrees of the vertices.  (Note that there is no factor
of 2 in the denominator now.)  Then the equivalent of
Eq.~\eqref{eq:moddef2} is
\begin{equation}
Q = {1\over m} \sum_{ij} \biggl[ A_{ij} - {\kin_i\kout_j\over m} \biggr]
    \delta_{c_i,c_j},
\label{eq:moddef3}
\end{equation}
where $A_{ij}$ is defined in the conventional manner to be 1 if there is an
edge from~$j$ to~$i$ and zero otherwise.  Note that indeed edges $j\to i$
make larger contributions to this expression if $\kin_i$ and/or $\kout_j$
is small.

Now we search for the division of the network into communities $\set{c_i}$
such that~$Q$ is maximized.  One can in principle make use of any of the
methods previously applied to modularity maximization, such as simulated
annealing or greedy algorithms.  Here we derive the appropriate
generalization of the spectral optimization method of
Newman~\cite{Newman2006b}, which is both computationally efficient and
appears to give excellent results in practice.

We consider first the simplified problem of dividing a directed network
into just two communities.  We define $s_i$ to be $+1$ if vertex~$i$ is
assigned to community~1 and $-1$ if it is assigned to community~2.  Note
that this implies that $\sum_i s_i^2=n$.  Then $\delta_{c_i,c_j} =
\frac12(s_is_j+1)$ and
\begin{align}
Q &= {1\over2m} \sum_{ij} \biggl[ A_{ij} - {\kin_i\kout_j\over m} \biggr]
    (s_is_j+1) \nonumber\\
  &= {1\over2m} \sum_{ij} s_i B_{ij} s_j
   = {1\over2m} \vec{s}^T\mat{B}\vec{s},
\label{eq:matq}
\end{align}
where $\vec{s}$ is the vector whose elements are the~$s_i$, $\mat{B}$~is
the so-called modularity matrix with elements
\begin{equation}
B_{ij} = A_{ij} - {\kin_i\kout_j\over m},
\label{eq:Bmatrix}
\end{equation}
and we have made use of $\sum_{ij} A_{ij} = \sum_i \kin_i = \sum_j \kout_j
= m$.  Our goal is now to find the~$\vec{s}$ that maximizes~$Q$ for a
given~$\mat{B}$.

In the undirected case the modularity matrix is symmetric but in the
present case it is, in general, not, and the lack of symmetry will cause
technical problems if we blindly attempt to duplicate the eigenvector-based
machinery presented for undirected networks in~\cite{Newman2006b}.
Luckily, however, we can easily restore symmetry to our problem by
adding~\eqref{eq:matq} to its own transpose to give
\begin{equation}
Q = {1\over4m} \vec{s}^T(\mat{B}+\mat{B}^T)\vec{s}.
\label{eq:matqsym}
\end{equation}
The matrix $\mat{B}+\mat{B}^T$ is now manifestly symmetric and it is on
this symmetric matrix that we focus forthwith.  Notice that
$\mat{B}+\mat{B}^T$ is not the same as the modularity matrix for a
symmetrized version of the network in which direction is ignored and hence
we expect methods based on the true directed modularity to give different
results, in general, to methods based on the undirected version.

The leading constant $1/4m$ in Eq.~\eqref{eq:matqsym} is conventional, but
makes no difference to the position of the maximum of~$Q$, so for the sake
of clarity we neglect it in defining our optimization procedure.

Following~\cite{Newman2006b}, we now write $\vec{s}$ as a linear
combination of the eigenvectors~$\vec{v}_i$ of $\mat{B}+\mat{B}^T$ thus:
$\vec{s} = \sum_i a_i \vec{v}_i$ with $a_i = \vec{v}_i^T\cdot\vec{s}$.
Then
\begin{equation}
Q = \sum_i a_i\vec{v}_i^T (\mat{B}+\mat{B}^T) \sum_j a_j \vec{v}_j
  = \sum_i \beta_i (\vec{v}_i^T\cdot\vec{s})^2,
\label{eq:modularitySym2}
\end{equation}
where $\beta_i$ is the eigenvalue of $\mat{B}+\mat{B}^T$ corresponding to
eigenvector~$\vec{v}_i$.  Let us assume the eigenvalues to be labeled in
decreasing order $\beta_1 \geq \beta_2 \geq \ldots \geq \beta_n$.  Under
the normalization constraint $\vec{s}^T\cdot\vec{s}=n$ the maximum of~$Q$
is achieved when $\vec{s}$ is parallel to the leading
eigenvector~$\vec{v}_1$, but normally this solution is forbidden by the
additional condition that $s_i=\pm1$.  We do the best we can, however, and
make $\vec{s}$ as close as possible to parallel with~$\vec{v}_1$, meaning
we choose the value of~$\vec{s}$ that maximizes $\vec{v}_1^T\cdot\vec{s}$.
It is straightforward to show that this gives $s_i=+1$ if $v_i^{(1)}>0$ and
$s_i=-1$ if $v_i^{(1)}<0$, where $v_i^{(1)}$ is the $i$th element
of~$\vec{v}_1$.  (If $v_i^{(1)}=0$ then $s_i=\pm1$ are equally good
solutions to the maximization problem.)

Thus we arrive at a simple algorithm for splitting a network: we calculate
the eigenvector corresponding to the largest positive eigenvalue of the
symmetric matrix $\mat{B}+\mat{B}^T$ and then assign communities based on
the signs of the elements of the eigenvector.

As in the undirected case, the spectral method typically provides an
excellent guide to the broad outlines of the optimal partition, but may err
in the case of individual vertices, a situation that can be remedied by
adding a ``fine-tuning'' stage to the algorithm in which vertices are moved
back and forth between communities in an effort to increase the modularity,
until no further improvements can be made~\cite{Newman2006b}.  We have
incorporated such a fine-tuning in all the calculations presented here.

So far we have discussed the division of a network into two communities.
There are a variety of ways of generalizing the approach to more than two
communities but the simplest, which we adopt here, is repeated bisection.
That is, we first divide the network into two groups using the algorithm
above and then divide those groups and so forth.  The process stops when we
reach a point at which further division does not increase the total
modularity of the network.

The subdivision of a community contained within a larger network requires a
slight generalization of the method above.  Consider the change in
modularity $\Delta Q$ of an entire network when a community~$g$ within it
is subdivided and, defining $s_i$ as before for vertices in~$g$, we find
\begin{align}
\Delta Q &= {1\over2m} \Biggl[ \sum_{i,j \in g} (B_{ij} + B_{ji})
            {s_i s_j + 1\over2}
            - \sum_{i,j \in g} (B_{ij} + B_{ji})\Biggr] \nonumber\\
         &= {1\over4m} \sum_{i,j \in g} \biggl[ (B_{ij} + B_{ji})
            - \delta_{ij} \sum_{k \in g} (B_{ik} + B_{ki}) \biggr] s_i s_j
            \nonumber\\
         &= {1\over4m} \vec{s}^T
            \Bigl(\mat{B}^{(g)} + {\mat{B}^{(g)}}^T \Bigr) \vec{s}
\label{eq:deltaQ}
\end{align}
where we have made use of $s_i^2=1$ for all~$i$ and
\begin{equation}
B_{ij}^{(g)} = B_{ij} - \delta_{ij} \sum_{k\in g} B_{ik}.
\label{eq:genbmatrix}
\end{equation}
In other words, $\mat{B}^{(g)}$~is the submatrix of $\mat{B}$ for the
subgraph~$g$ with the sum of each row subtracted from the corresponding
diagonal element.  Although $\mat{B}^{(g)}$, like~$\mat{B}$, is in general
asymmetric, the sum $\mat{B}^{(g)} + {\mat{B}^{(g)}}^T$ is symmetric and
hence Eq.~\eqref{eq:deltaQ} has the same functional form as
Eq.~\eqref{eq:matqsym} and we can apply the same method to maximize~$\Delta
Q$.

Our complete algorithm for discovering communities or groups in a directed
network is thus as follows.  We construct the modularity matrix,
Eq.~\eqref{eq:Bmatrix}, for the network and find the most positive
eigenvalue of the symmetric matrix $\mat{B} + \mat{B}^T$ and the
corresponding eigenvector.  Each vertex is assigned to one of two groups
depending on the sign of the corresponding element of the eigenvector and
then we fine-tune the assignments as described above to maximize the
modularity.  We then further subdivide the communities using the same
method, but with the generalized modularity matrix,
Eq.~\eqref{eq:genbmatrix}, fine tuning after each division.  If the
algorithm finds no division giving a positive value of $\Delta Q$ for a
particular community then we can increase the modularity no further by
subdividing this community and we leave it alone.  When all communities
reach this state the algorithm ends.

We now give a number of examples of the application of our method.  We
consider four different directed networks of varying degrees of complexity,
starting with a relatively simple but important example: the world wide
web.

Weblogs or ``blogs'' are personal web sites on which their proprietors
record brief thoughts on topics of their choosing, often with links to
other blogs with related discussions.  In a recent study, Adamic and
Glance~\cite{AG05} looked at a network of 1225 blogs focusing on US
politics.  In this network the vertices represent the blogs and there is a
directed link between vertices if one blog links to another.  Adamic and
Glance also characterized the political persuasion of each blog as
conservative or liberal based on textual content.

When fed into our community finding algorithm, the blog network divides
into two clear communities, with one being composed almost entirely of
conservative blogs and the other of liberal blogs.  (The algorithm places
97\% of the blogs characterized by Adamic and Glance as conservative in the
first community and 93\% of those characterized as liberal in the second.)
The algorithm finds no subdivision of either community that gives any
increase in the modularity, indicating that the network consists of only
two tightly knit communities corresponding closely to the traditional
left-right division of US politics.  This serves as a particularly clear
demonstration of the algorithm's ability to find meaningful structure in
network data.  But on the other hand this particular network gives very
similar results when analyzed using the undirected form of the spectral
modularity algorithm, in which edge direction is entirely
ignored~\cite{Newman2006b}.  The principal interest in our algorithm
derives from its ability to find structure in networks where the simpler
undirected version fails, so let us turn to examples of this kind.

\begin{figure}
\begin{center}
\hfill
\subfigure[]{\includegraphics[width=3.75cm,angle=90]{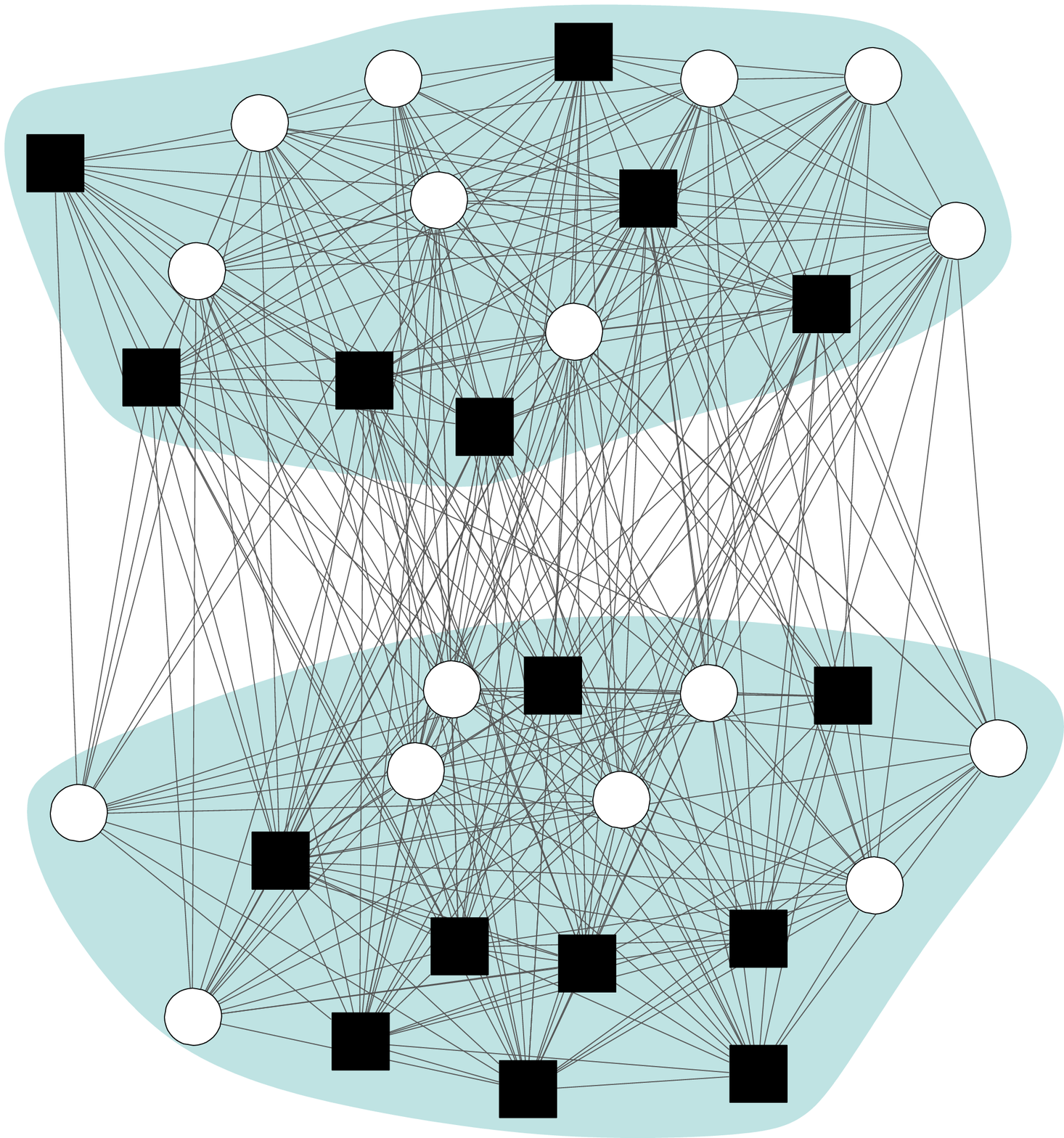}}
\hfill
\subfigure[]{\includegraphics[width=3.75cm,angle=90]{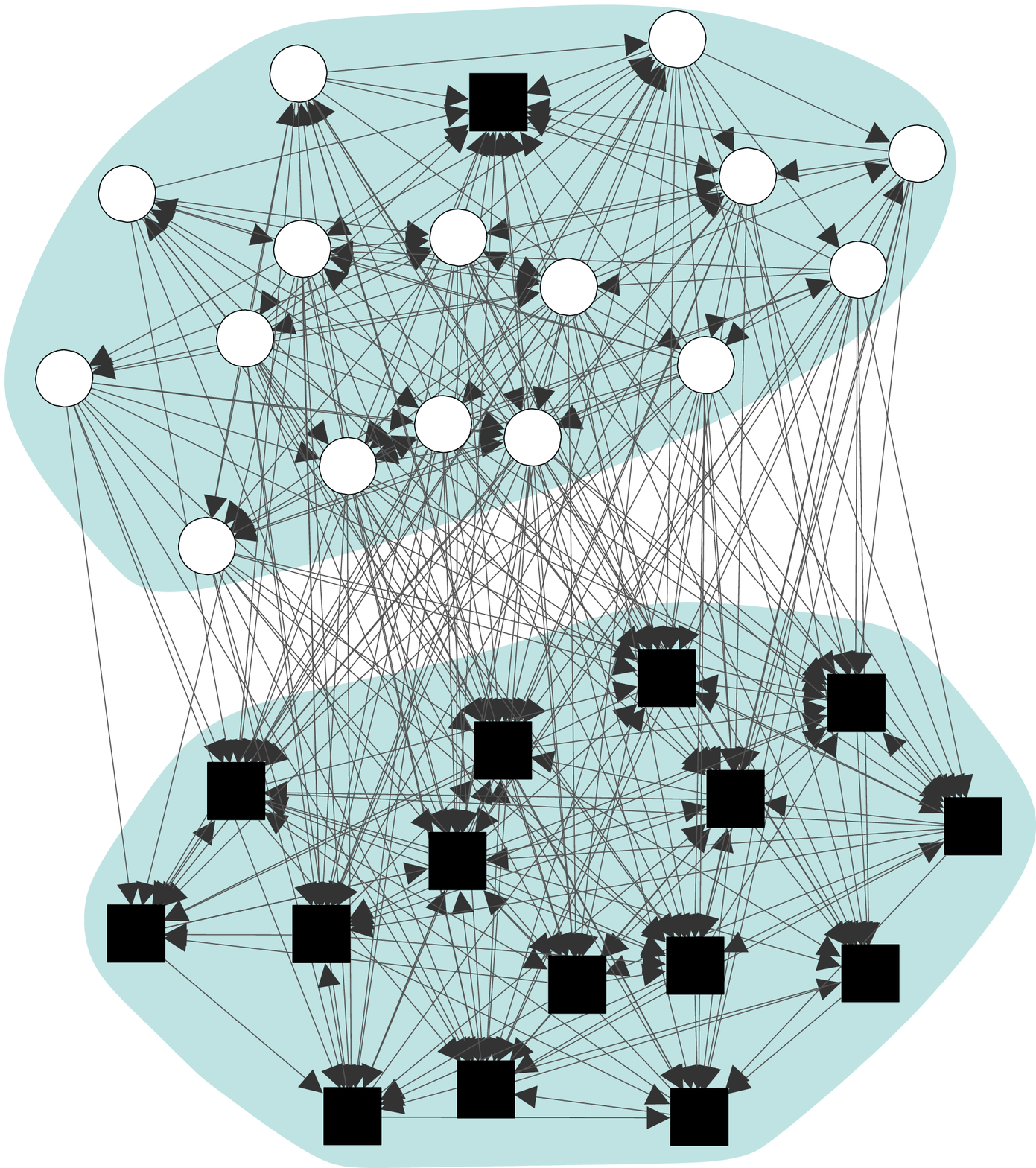}}
\hfill\null
\end{center}
\caption{Community assignments for the two-community random network
  described in the text from (a)~a standard undirected modularity
  maximization which ignores edge direction and (b)~the algorithm of this
  paper.  The shaded regions represent the communities discovered by the
  algorithms.  The true community assignments are denoted by vertex shape.}
\label{fig:twoGroupsExample}
\end{figure}

For illustrative purposes, we first consider an artificial
computer-generated network, designed specifically to test the performance
of the algorithm.  In this network of 32 vertices, vertex pairs are
connected by edges independently and uniformly at random with some
probability~$p$.  The edges are initially undirected.  The network is then
divided into two groups of 16 vertices each and edges that fall within
groups are assigned directions at random but edges between groups are
biased so that they are more likely to point from group 1 to group 2 than
\textit{vice versa}.

By construction, there is no community structure to be found in this
network if we ignore edge directions---the positions of the edges are
entirely random---and this is confirmed in
Fig.~\ref{fig:twoGroupsExample}a, which shows the results of the
application of the undirected modularity maximization algorithm.  If we
take the directions into account, however, using the algorithm presented in
this paper, the two communities are detected almost perfectly: just one
vertex out of 32 is misclassified---see Fig.~\ref{fig:twoGroupsExample}b.

\begin{figure}
\begin{center}
\hfill
\subfigure[]{\includegraphics[width=4cm,angle=90]{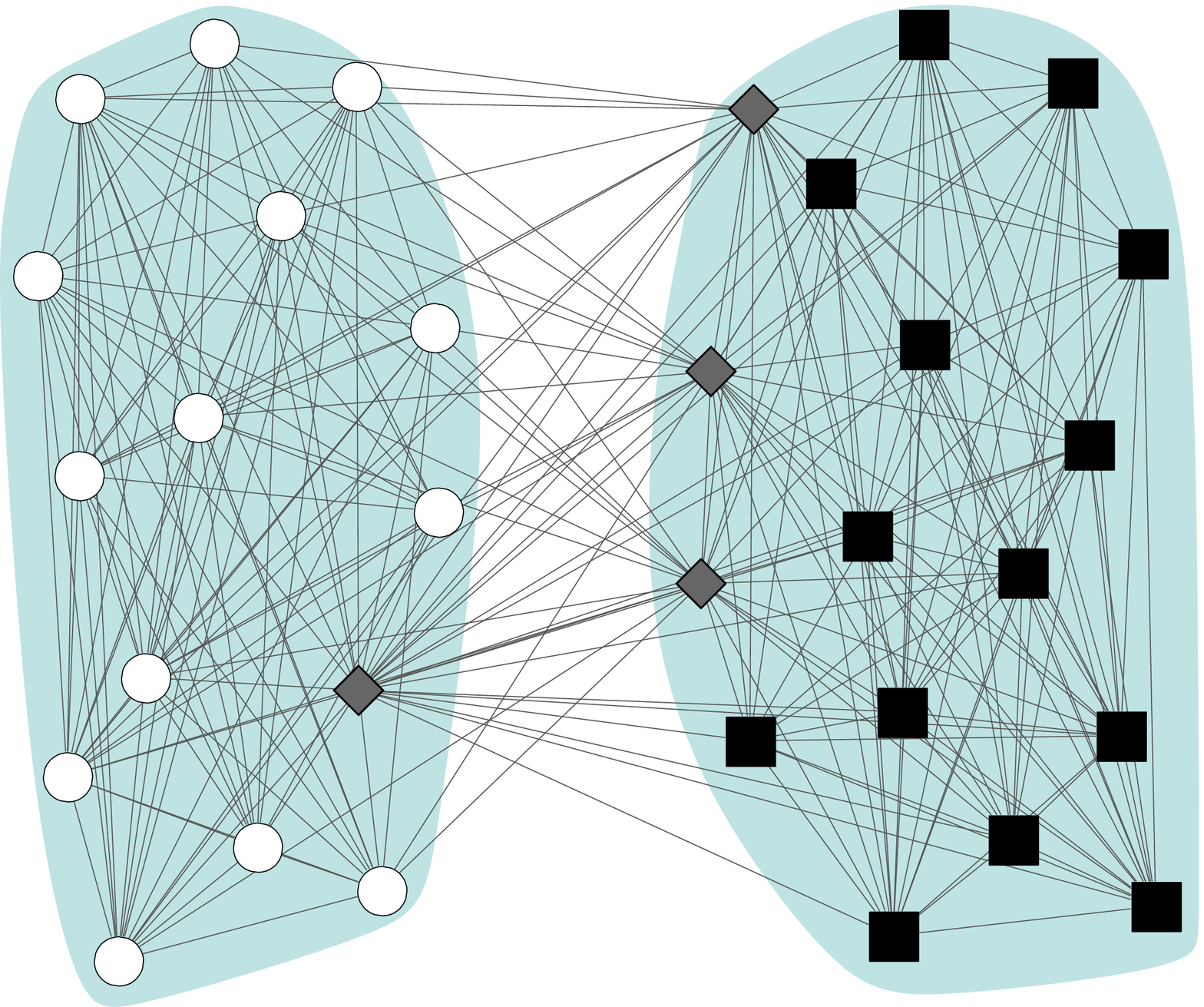}}
\hfill
\subfigure[]{\includegraphics[width=4cm,angle=90]{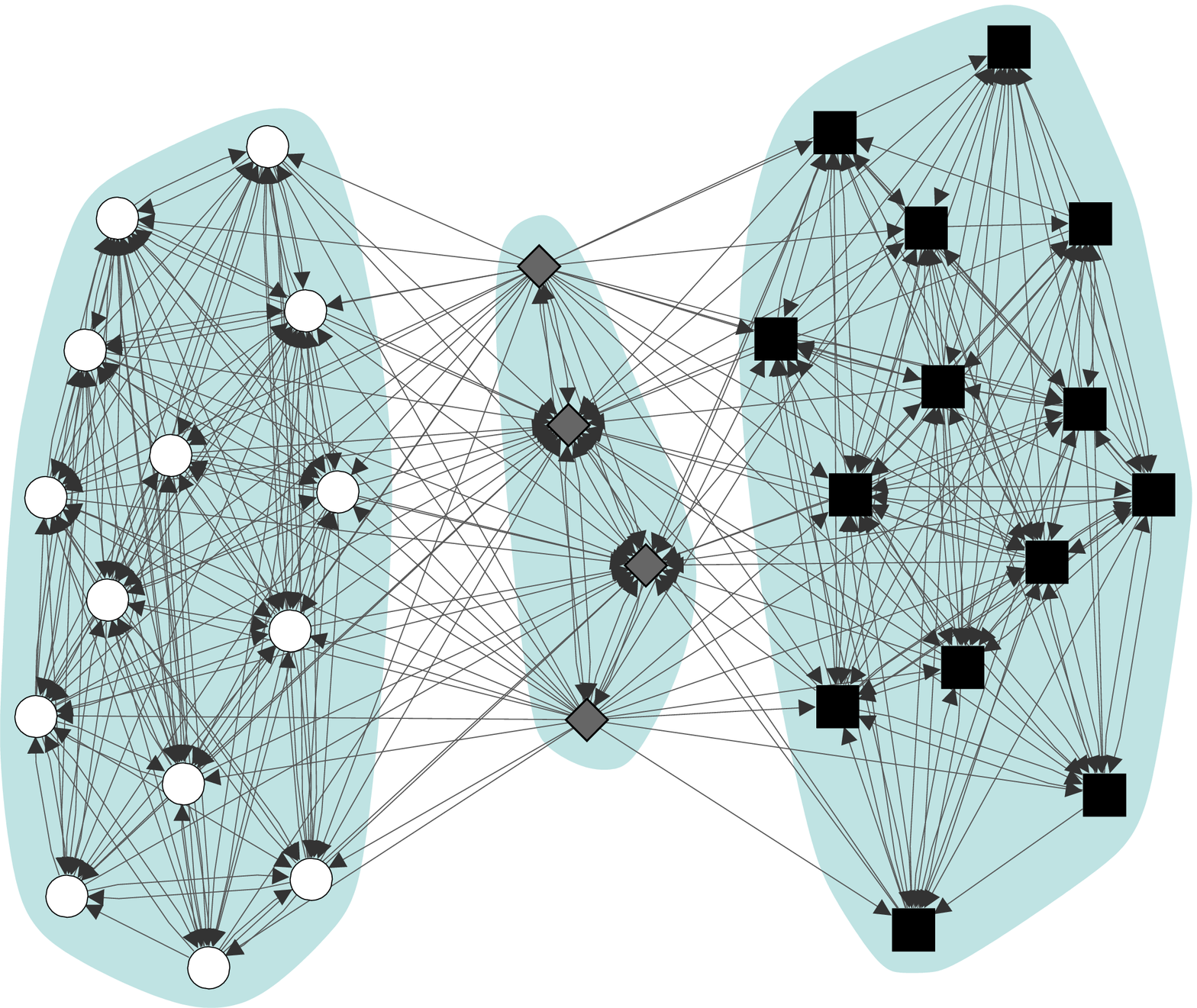}}
\hfill\null
\end{center}
\caption{Community assignments for the three-community random network
  described in the text as generated by (a)~standard undirected modularity
  maximization and (b)~the algorithm of this paper.}
\label{fig:haExample}
\end{figure}

Even in networks where there is clear community structure contained in the
positions of the edges it is still possible for the directions to contain
additional useful information.  As an example of this type of behavior,
consider the network shown in Fig.~\ref{fig:haExample}, which has 32
vertices and three communities.  For two of the communities, containing 14
vertices each, there is a high probability of connection between pairs of
vertices that fall in the same community but a lower probability if one of
the vertices is in a different community.  Structure of this kind, in which
edge direction does not play a role, can in principle be found by
algorithms designed for undirected networks.  The third community, however,
is different.  It has four vertices, each of which has a high probability
of connection to every other vertex.  The only feature that distinguishes
this third community as separate is the direction of its edges---two of the
four vertices have high probability of ingoing edges, the other two have
high probability of outgoing edges, and there are also a small number of
additional edges running from the former to the latter.  These last edges
are statistically surprising in the sense considered here and hence tend to
bind the third community together.

Applied to this network, the standard undirected community detection
algorithm finds the two large communities with ease, but the remaining
community is not found and its vertices are dispersed by the algorithm
among the other communities (Fig.~\ref{fig:haExample}a).  Our directed
algorithm, on the other hand, finds all three communities without
difficulty (Fig.~\ref{fig:haExample}b).  Again the algorithm has made use
of information contained in the edge directions to identify community
structures not accessible to previous methods.

\begin{figure}
\begin{center}
\includegraphics[width=8cm]{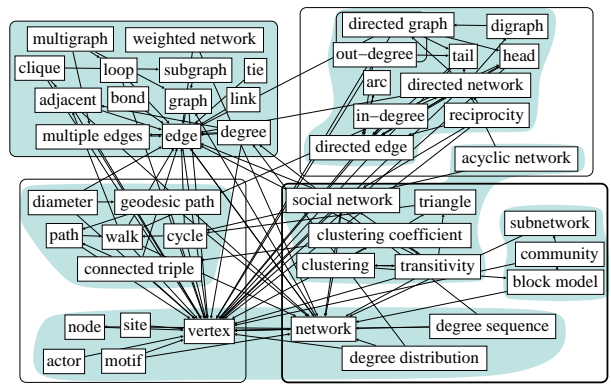}
\end{center}
\caption{The network of technical terms described in the text along with
  the community assignments determined by a standard undirected modularity
  maximization (boxes) and the algorithm of this paper (shaded groups).}
\label{fig:networkWords}
\end{figure}

Returning now to real-world networks, we show in
Fig.~\ref{fig:networkWords} a further example of the performance of our
algorithm on, in this case, a word network.  The network represents
connections between a set of technical terms, such as ``vertex'' and
``edge,'' contained in a glossary of network jargon derived from recent
review papers by Newman~\cite{Newman2003c} and
Boccaletti~\etal~\cite{Boccaletti2006}.  Vertices in this network represent
terms and there is a directed edge from one vertex to another if the first
glossary term was used in the definition of the second.  Because circular
definitions are unhelpful and normally avoided, most edges in the network
are not reciprocated.

Figure~\ref{fig:networkWords} shows the communities found in this network
by our directed modularity algorithm.  The algorithm finds six communities
in this case that appear to correspond to groupings of terms clustered
around a few basic concepts.  For instance, one group deals with words
describing basic network structure, such as ``edge'' and ``graph,'' while
another deals with terms describing directed networks.  A third group
contains the terms ``vertex'' and ``degree'' and related concepts and the
remaining groups are associated with clustering, communities, and paths
respectively.  Thus, the algorithm again appears to find meaningful
structure in the network, of the type that could be useful in understanding
the broader shape of otherwise poorly understood systems.

We have also applied the undirected modularity maximization algorithm to
this same network, which results in four groups.  Two of these are closely
similar to ones found by the directed algorithm---the groups dealing with
edges and with directed networks.  The other groups, however, contain a mix
of terms that do not correspond closely to any obvious network concepts,
with words like ``vertex,'' ``diameter,'' ``cycle,'' and ``motif'' grouped
together.  As discussed above, the undirected algorithm has less
information at its disposal, the directions of the edges having been
discarded, so it is natural that it is unable to detect some of the structure
found by its directed counterpart.

In summary, we have presented a method for detecting community structure in
directed networks that makes explicit use of information contained in edge
directions, information that most other algorithms discard.  Our method is
an extension of the established modularity maximization method widely used
to determine community structure in undirected networks.  We have applied
the method to a variety of networks, both real and simulated, showing that
it is able to recover known community structure in previously studied
networks and extract additional and revealing structural information not
available to algorithms that ignore edge direction.  The computational
efficiency of the algorithm is essentially identical to that of the
corresponding algorithm for undirected networks and hence we see no reason
to continue to use the undirected algorithm on directed graphs; we
recommend the use of the full directed algorithm in all cases where
researchers wish to analyze both edge placement and edge direction.

The authors thank Luis Amaral, Roger Guimer\`a, and Marta Sales-Pardo for
useful conversations.  This work was funded in part by the National Science
Foundation under grant DMS--0405348 and by the James S. McDonnell
Foundation.

\end{document}